\RequirePackage{amsmath,amssymb}
\documentclass{llncs}
\usepackage{makeidx}  % allows for indexgeneration
\usepackage{graphicx}
\begin{document}
\frontmatter          % for the preliminaries
\pagestyle{headings}  % switches on printing of running heads
%\addtocmark{Hamiltonian Mechanics} % additional mark in the TOC
%

%
%
%
%

%

%
%\tableofcontents
%
\mainmatter              % start of the contributions
\title{Protecting suppliers’ private information: the case of stock levels and the impact of correlated items}
\titlerunning{Protecting suppliers' information}  % abbreviated title (for running head)
%                                     also used for the TOC unless
%                                     \toctitle is used
%
\author{Maurizio Naldi\inst{1} \and Giuseppe D'Acquisto\inst{2}
}
\authorrunning{Naldi and D'Acquisto} % abbreviated author list (for running head)
%
%%%% list of authors for the TOC (use if author list has to be modified)
\tocauthor{Maurizio Naldi and Giuseppe D'Acquisto}
\institute{University of Rome at Tor Vergata, Rome 00133, Italy,\\
\email{naldi@disp.uniroma2.it}
\and
University of Rome at Tor Vergata, Rome 00133, Italy,\\
\email{dacquisto@ing.uniroma2.it}}

\maketitle              % typeset the title of the contribution

\begin{abstract}
A marketplace is defined where the private data of suppliers (e.g., prosumers) are protected, so that neither their identity nor their level of stock is made known to end customers, while they can sell their products at a reduced price. A broker acts as an intermediary, which takes care of providing the items missing to meet the customers' demand and allows end customers to take advantages of reduced prices through the subscription of option contracts. Formulas are provided for the option price under three different probability models for the availability of items. Option pricing allows the broker to partially transfer its risk on end customers.
\keywords{privacy, statistical databases, supply chain}
\end{abstract}
\section{Introduction}
Though privacy has been defined as \textit{the claim of individuals, groups, or institutions to determine for themselves when, how, and to what extent information about them is communicated to others} \cite{westin1968privacy}, the right of individuals, rather than companies, to protect their personal data has so far been the focus of most privacy studies. This is especially true in the context of marketplaces, where consumers may release, deliberately or not,  details about themselves and the items they purchase. Algorithms and platforms have been devised to enforce customers' privacy requirements  (see, e.g., \cite{karjoth2003platform,kalvenes2006design}), but little attention has been paid to the other side of trading, i.e., companies selling their products.

However, a company may wish to select the level of information it provides to its customers, but the widespread adoption of e-shops divulges a lot of details about company's operations, not just to prospective customers but to everyone accessing the e-commerce platform, including competitors. For example, companies may wish to keep their level of stock for a given product secret. The issue is even more delicate when traders are actually prosumers, who, acting as sellers, may reveal personal data; an example of such a context is that of smart grids \cite{clements2010cyber}. Generally speaking, individuals acting as suppliers may wish not to be profiled and rather keep secret the products they happen to own. The definition of a marketplace where suppliers can sell their products while retaining privacy is then a relevant issue.

Such a definition has been proposed in \cite{MNPrinf15}, where the information to be protected is the identity of the sellers and their level of stock, and the role of a broker is envisaged. It was shown that such a marketplace may be set up with benefits for all the stakeholders (a broker/producer, privacy-aware suppliers, and end customers) through the use of differential privacy mechanisms \cite{dwork2011} and option contracts subscribed by end customers \cite{MNcns15}. A general formula for option pricing has been derived in \cite{MNcns15}, where nothing at all is known about the actual availability of items by the suppliers.

In this paper, we embrace the marketplace definition provided in \cite{MNPrinf15} and consider the case where the broker, though it doesn't know the actual number of available items, may adopt a hypothesis concerning their probability distribution. In particular, we consider the cases where the stocks owned by suppliers exhibits either full or null correlation (i.e., respectively perfectly correlated stocks and independent suppliers) and a third case where a uniform distribution applies (representing a mild correlation). For each case we derive a formula for the option price. We show that the using such an option price allows the broker to transfer its risk to end customers.

The paper is organized as follows. In Section \ref{market} we define the marketplace and the stakeholders, while the option contract between the broker and  end customers is described in Section \ref{opzioni}. In Section \ref{ava} we describe the three models for the availability of items, which are used to derive the option pricing formulas in Section \ref{formulas}.

\section{A market for privacy-aware suppliers}
\label{market}
Let's consider a database of suppliers where information can be obtained about the availability of a set of items, but suppliers are somewhat screened. Suppliers could be vendors whose typical line of business does not include those products or who wish to get rid of some remainders, or individuals (prosumers) who happen to have those products in their availability. For example, the database could contain the number of items available for sale at each supplier, so that the vertical sum across all suppliers included in the database would tell us the overall number of items available. Such a database, providing statistics about the entities included in it, is called a statistical database \cite{shoshani1982}. However, in a statistical database, releasing statistical information may compromise the privacy of individual contributors. But suppliers may wish not to divulge those information; for example they do not want competitors (who could access the database) to know their level of stock, or, as individuals, they do not wish to be profiled about the items they own. If suppliers wish to be screened, a curator may sit between the users, posing the query, and the database. The responses to these queries may be modified by the curator in order to protect the privacy of the contributors \cite{dwork2008}, for example so as not to tell us exactly either which supplier can provide us with those items or how many items in the set are available. Instead of providing the exact number, the database provides us with an obfuscated number, which is more or less close to the exact figure. A mechanism to achieve differential privacy is the use of noisy sums: the response to a counting query is the sum of the true figure and some noise \cite{dwork2011}. The use of a statistical database plus the use of noisy sums may therefore protect the private information of suppliers.

When end customers demand for a number of items, the uncertainty surrounding the actual availability of those items doesn't allow  to close deals. In the presence of such privacy constraints, we postulate that a market can develop through the introduction of a broker/producer and the use of option contracts.
    
Let's consider the case where end customers demand for $k^{*}$ items. A broker commits to provide them with the number of items required. In fact, the broker may procure those items either by producing them itself (at a unit production cost $c_{p}$) or by resorting to \textit{privacy-aware suppliers}, whose availability is known through the statistical database previously mentioned. As already said, privacy-aware suppliers do not release full information about the availability of their products, but release instead an obfuscated number $\hat{k}$, which is generally different from the true number $k$ of items that they can provide (though the broker may obtain a refined estimate of the true number through Bayesian analysis \cite{naldi2014differential}).  

The privacy enjoyed by privacy-aware suppliers is reflected in the price $c_{s}$ they advertise. Prices set by privacy-aware suppliers depend on the level of obfuscation (i.e. privacy protection): the higher the level of obfuscation (embodied by the variance of the added noise), the lower the price. Assuming $c_{s}<c_{p}$, the broker has a real advantage to procure as many items as it can through privacy-aware suppliers at the reduced price $c_{s}$, and transfer part of that benefit to end customers by setting a lower end price. If the availability of items is not enough to satisfy the demand ($k<k^{*}$), the broker/producer produces the remaining items (but does not enjoy the full benefit of the reduced price).

\begin{figure}[htbp]
\begin{center}	
  \includegraphics[width=.85\columnwidth]{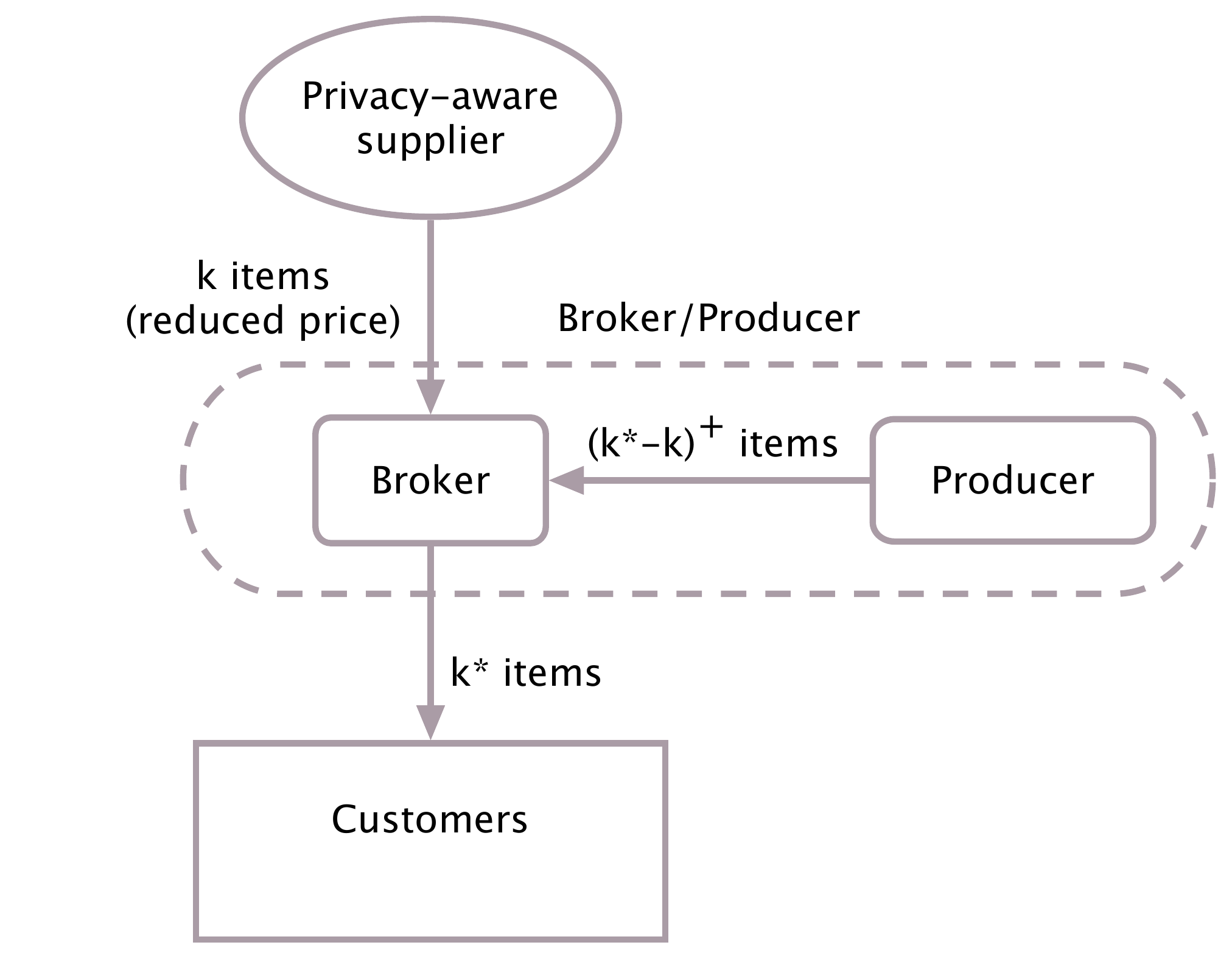}
	\caption{Supply relationships}
	\label{fig:sup}
\end{center}
\end{figure}

In order to exploit the offer by privacy-aware suppliers, the broker submits a query to the statistical database containing information about the availability of items and pays a fixed amount $c_{q}$ and receives the noisy response $\hat{k}$.  It commits to buy all the $k$ items available, though they may exceed the actual demand $k^{*}$. When the actual number of available items is disclosed (at delivery), it pays the privacy-aware suppliers the overall amount $c_{s}k$. If the demand is fully met ($k>k^{*}$) the broker does not have to produce any item; otherwise, the broker has to produce $k^{*}-k$ items at the unit cost $c_{p}$. The resulting supply chain is shown in \figurename~\ref{fig:sup}. 

It is to be noted that both curator and the broker in the end know the exact number of items available by the suppliers, but they have (different) reasons to keep it private. In fact, the database curator does not have a direct contact with end customers and is not in the business of retailing. Instead, the broker's business relies on the privacy of those data for its intermediary role. In addition, the roles of the curator and the broker may have to be kept separate due to regulatory constraints.

However, such a procedure is not free of risks for the broker/producer, which, on the one hand commits to provide its customers with the required items, but on the other hand is subject to the uncertainty determined by the unknown availability of items delivered by privacy-aware suppliers, with the risks deriving from the commitments to buy all the items available and, if required, to produce the remaining ones at a higher cost.

The broker/producer has therefore to hedge against such risks. A way we suggest is to resort to option contracts, which are described in the next section.

\section{Risk coverage through options}
\label{opzioni}
As seen in the previous section, the broker/producer undergoes a risk when resorting to privacy-aware suppliers in order to meet customers' demand at a reduced cost, a benefit which it transfers to end customers through a reduced price. It needs however to hedge against such a risk. In this section we describe a mechanism, based on option contracts, by which it can achieve such protection.

Since the stakeholders that ultimately benefit from resorting to privacy-aware suppliers are end customers, the broker/producer may transfer some of that risk to them, asking them to pay a price to get the right to buy the desired number of items at a predetermined lower price  (i.e., a booking fee). In the language of financial markets, this is a \textit{call} option, since it endows the end customer with the right to buy \cite{hull2006options}. End customers are then required to subscribe a call option to be sure to get the right number of items they wish at a reduced price.

A critical issue in option contract is setting the right price. In the typical scenario, the amount to be paid for the option contract is expected to depend on the current value of the items for sale, the predetermined price to be paid if the option is exercised, and the expected behaviour of the item's value in the period from the option contract underwriting to the exercise time. A simple form of pricing is given by the Black-Scholes formula \cite{davis2010}, but a form tailored for the context is to be derived here.

The general expression for the option price is
\begin{equation}
\label{genprice}
c_{\textrm{opt}} =  \mathbb{E}[c_{\textrm{s}}(k-k^{*})^{+} \vert \hat{k}],
\end{equation}
which considers the risk due to the extra-cost of buying the excess items provided by privacy-aware suppliers. In  \cite{MNcns15}, the risk of having to buy the items exceeding the demand has been analysed, and the following pricing formula has been derived for the case where Laplacian noise is added to form the noisy sum \cite{Sarathy2011}: 
\begin{equation}
\label{optprice}
c_{\textrm{opt}} =  c_{\textrm{s}}\left[(\hat{k}-k^{*})^{+}+\frac{1}{2\lambda}e^{-\lambda \vert \hat{k} - k^{*}\vert}\right],
\end{equation}
where $\lambda$ is the shape parameter of the Laplace distribution: the smaller $\lambda$, the greater the differential privacy. In that formula, the declared value $\hat{k}$ was considered as the best estimate of the actual availability.

However, if we have some \textit{a priori} information about the actual availability, we can gain a better estimate for it and obtain a more accurate formula for the option price. In \cite{naldi2014differential}, we have shown that Bayesian analysis mat be used to obtain a better estimate of the figure previously obfuscated through the addition of Laplace noise. Here we exploit the same mechanism to obtain a better estimate of the actual availability. In the next section we describe three models that may provide the \textit{a priori} information we need to apply Bayes estimation.

\section{Models for items availability}
\label{ava}
The extra-cost incurred by the broker to get the items through privacy-aware suppliers depends on the number of items actually available. This is unknown to the broker till the disclosure by those suppliers when setting the deal. However, models may be adopted for the envisaged availability that allow us to obtain a formula for the option price once the declared availability $\hat{k}$ is made known. In this section, we describe three models, which represent three paradigmatic situations: unit correlation, independent suppliers, and uniform distribution (which represents a mild correlation).

The case of unit correlation applies when either all the suppliers have the item or none of them have it. Their behaviour shows therefore full correlation, hence the name given to the model. The number of availabile items may therefore take just either of two values: 0 or $n$. The probability associated to the two cases is
\begin{equation}
\label{probunit}
 \begin{split}
 \mathbb{P}[k=0] &= 1-p\\
 \mathbb{P}[k=n] &= p.
 \end{split}
 \end{equation} 

The case opposite to full correlation is that of no correlation at all, where the availability of the item by any supplier is independent of any other suppliers. If we now indicate the probability of any supplier to have the item by $p$, the probability of the number of available items follows a binomial distribution with parameters $p$ and $n$:
\begin{equation}
\mathbb{P}[k=i] = \binom{n}{i} p^{i}(1-p)^{n-i} \qquad i=0,1,\ldots ,n. 
\end{equation}

Finally, as an intermediate case between those of no correlation and unit correlation, we can consider the case of uniform distribution
\begin{equation}
\mathbb{P}[k=i] = \frac{1}{n+1} \qquad i=0,1,\ldots ,n.
\end{equation}

\section{Formulas for risk transfer}
\label{formulas}
After defining the three paradigmatic models for the availability of items, in this section we derive the pricing formulas for the three cases.

\subsection{Unit correlation}
In the case of unit correlation, the number of actually available items may take either of two values $k=0$ and $k=n$, as defined in Equation (\ref{probunit}). 

The extra-cost is then
\begin{equation}
\label{eunicorr}
\begin{split}
\mathbb{E}[c_{\textrm{s}}(k-k^{*})^{+}\vert \hat{k}=x] &= c_{\textrm{s}}\{ \mathbb{P}[k=0\vert \hat{k}=x](0-k^{*})^{+} + \mathbb{P}[k=n\vert \hat{k}=x](n-k^{*})^{+}\}\\
&= c_{\textrm{s}}\mathbb{P}[k=n\vert \hat{k}=x](n-k^{*})\\
\end{split}
\end{equation}

This expression is still dependent on the conditional probability $\mathbb{P}[k=n\vert \hat{k}=x]$, which we may obtain through Bayes' theorem
\begin{equation}
\label{bayesunicor}
\begin{split}
\mathbb{P}[k=n\vert \hat{k}=x] &= \frac{\mathbb{P}[k=n]\mathbb{P}[\hat{k}=x\vert k=n]}{\mathbb{P}[\hat{k}=x]}\\
&= \frac{\mathbb{P}[k=n]\mathbb{P}[\hat{k}=x\vert k=n]}{\mathbb{P}[\hat{k}=0]\vert\mathbb{P}[\hat{k}=x\vert k=0] + \mathbb{P}[\hat{k}=n]\vert\mathbb{P}[\hat{k}=x\vert k=n]}\\
&=\frac{p\frac{\lambda}{2}e^{-\lambda \vert n-x \vert}}{(1-p)\frac{\lambda}{2}e^{-\lambda \vert x \vert} + p\frac{\lambda}{2}e^{-\lambda \vert n-x \vert} }\\
&=\frac{1}{1+\frac{1-p}{p}e^{-\lambda[\vert x \vert - \vert n-x \vert]}}
\end{split}
\end{equation}
By replacing the expression (\ref{bayesunicor}) in the extra-cost expression (\ref{eunicorr}) we finally obtain
\begin{equation}
\label{eunicorrfin}
\mathbb{E}[c_{\textrm{s}}(k-k^{*})^{+}\vert \hat{k}=x] =  \frac{c_{\textrm{s}}(n-k^{*})}{1+\frac{1-p}{p}e^{-\lambda[\vert x \vert - \vert n-x \vert]}}
\end{equation}

If we assume that the declaration falls within the limits of the range of suppliers, i.e., $0\le \hat{k} \le n$, we have
\begin{equation}
\label{eunicorrfin2}
\begin{split}
c_{\textrm{opt}} &=  \frac{c_{\textrm{s}}(n-k^{*})}{1+\frac{1-p}{p}e^{-\lambda[ x  - (n-x) ]}}\\
&= \frac{c_{\textrm{s}}n(1-k^{*}/n)}{1+\frac{1-p}{p}e^{-\lambda (2x-n)}}
\end{split}
\end{equation}

If we consider the price as a function of the declared number of available items, we see that the price switches quite abruptly between two values. When $\hat{k}=0$, the normalized price is
\begin{equation}
\frac{c_{\textrm{opt}}}{c_{\textrm{s}}n}  = \frac{(1-k^{*}/n)}{1+\frac{1-p}{p}e^{\lambda n}} \simeq 0    \qquad n\gg 1
\end{equation}

Instead, when we are on the opposite side of declared values, $\hat{k} = n$, we have
\begin{equation}
\frac{c_{\textrm{opt}}}{c_{\textrm{s}}n}  = \frac{(1-k^{*}/n)}{1+\frac{1-p}{p}e^{-\lambda n}} \simeq  1-k^{*}/n   \qquad n\gg 1
\end{equation}
The turning point between the two values is $\hat{k}=n/2$

\begin{figure}
\begin{center}
\includegraphics[scale=0.4]{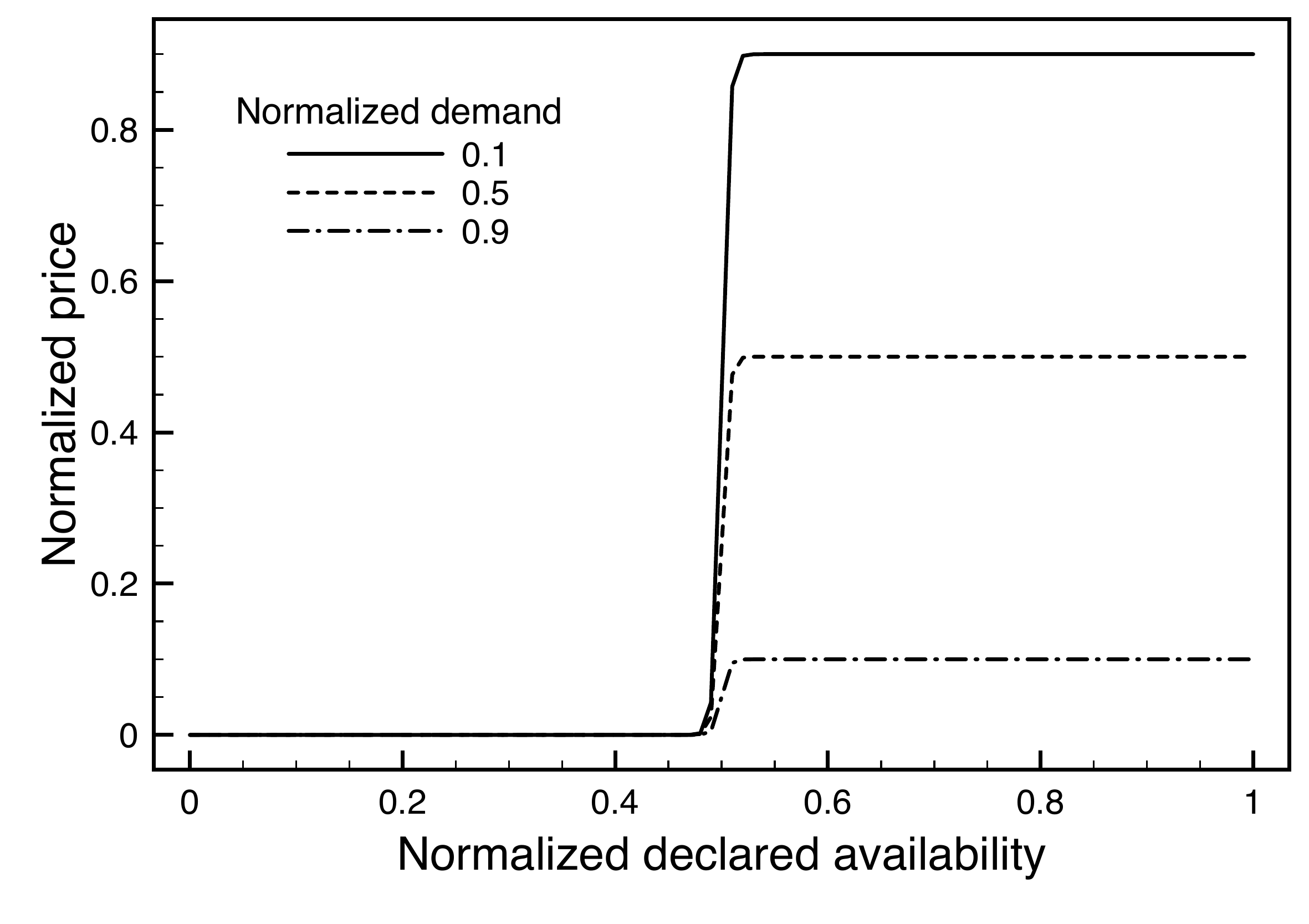} 
\caption{Impact of demand on option price (correlated suppliers)}
\label{fig:prixcorr}
\end{center}\end{figure}

A set of price curves are shown in \figurename~\ref{fig:prixcorr} for $n=100$ suppliers, $\lambda = 1.5$, and $p=0.5$. As we can see the normalized option price is practically zero when the declared availability is $\hat{k}<n/2$ and nearly $1-k^{*}/n$ when the declared availability is $\hat{k}>n/2$. The only parameters that actually impact the price are the demand $k^{*}$ and the declared availability $\hat{k}$. The probability $p$ of the suppliers having all the items plays a negligible role, just in the small range around $\hat{k}=n/2$. In practice, this means that the end customer switches from paying nothing (when the declared availability is less than half the number of suppliers) to paying almost the full price of the items available but not required (when the declared availability is more than half the number of suppliers). In the latter case there is practically a complete transfer of risk.

\subsection{Independent suppliers}
In the case of independent suppliers, each one having the item with probability $p$, the number of actually available items follows a binomial distribution with parameters $n$ and $p$, as described in Section \ref{ava}. 

Since the broker will incur an extra-cost for the items that it has to buy from the suppliers in excess of the actual demand, the extra-cost suffered by the broker is
\begin{equation}
\label{ind1}
c_{\textrm{opt}} =\mathbb{E}[c_{\textrm{s}}(k-k^{*})^{+}\vert x<\hat{k}<x+dx] = \sum_{i=k^{*}+1}^{n} c_{\textrm{s}}(i-k^{*})\mathbb{P}[k=i\vert x<\hat{k}<x+dx]
\end{equation} 

Again by Bayes' theorem we have
\begin{equation}
\begin{split}
\mathbb{P}[k=i \vert x<\hat{k}<x+dx] &= \frac{\mathbb{P}[k=i]\mathbb{P}[x<\hat{k}<x+dx \vert k=i]}{\mathbb{P}[x<\hat{k}<x+dx]}\\
&= \frac{\binom{n}{i} p^{i}(1-p)^{n-i}\frac{1}{2}\lambda e^{-\lambda \vert x-i \vert}}{\sum_{j=0}^{n}\binom{n}{j} p^{j}(1-p)^{n-j}\frac{1}{2}\lambda e^{-\lambda \vert x-j \vert}},
\end{split}
\end{equation}
that, when replaced in Equation (\ref{ind1}), provides
\begin{equation}
\label{ind2}
\begin{split}
c_{\textrm{opt}} &=  c_{\textrm{s}} \frac{\sum_{i=k^{*}+1}^{n}(i-k^{*})\binom{n}{i} p^{i}(1-p)^{n-i}\frac{1}{2}\lambda e^{-\lambda \vert x-i \vert}}{\sum_{j=0}^{n}\binom{n}{j} p^{j}(1-p)^{n-j}\frac{1}{2}\lambda e^{-\lambda \vert x-j \vert}}\\
&= c_{\textrm{s}} \frac{\sum_{i=k^{*}+1}^{n}(i-k^{*})\binom{n}{i} p^{i}(1-p)^{n-i} e^{-\lambda \vert x-i \vert}}{\sum_{j=0}^{n}\binom{n}{j} p^{j}(1-p)^{n-j} e^{-\lambda \vert x-j \vert}}
\end{split}
\end{equation} 

We now examine the dependence of the option price on the following parameters:
\begin{itemize}
\item Declared availability $\hat{k}$
\item Demand $k^{*}$
\item Probability $p$ of individual availability
\end{itemize}

\begin{figure}
\begin{center}
\includegraphics[scale=0.4]{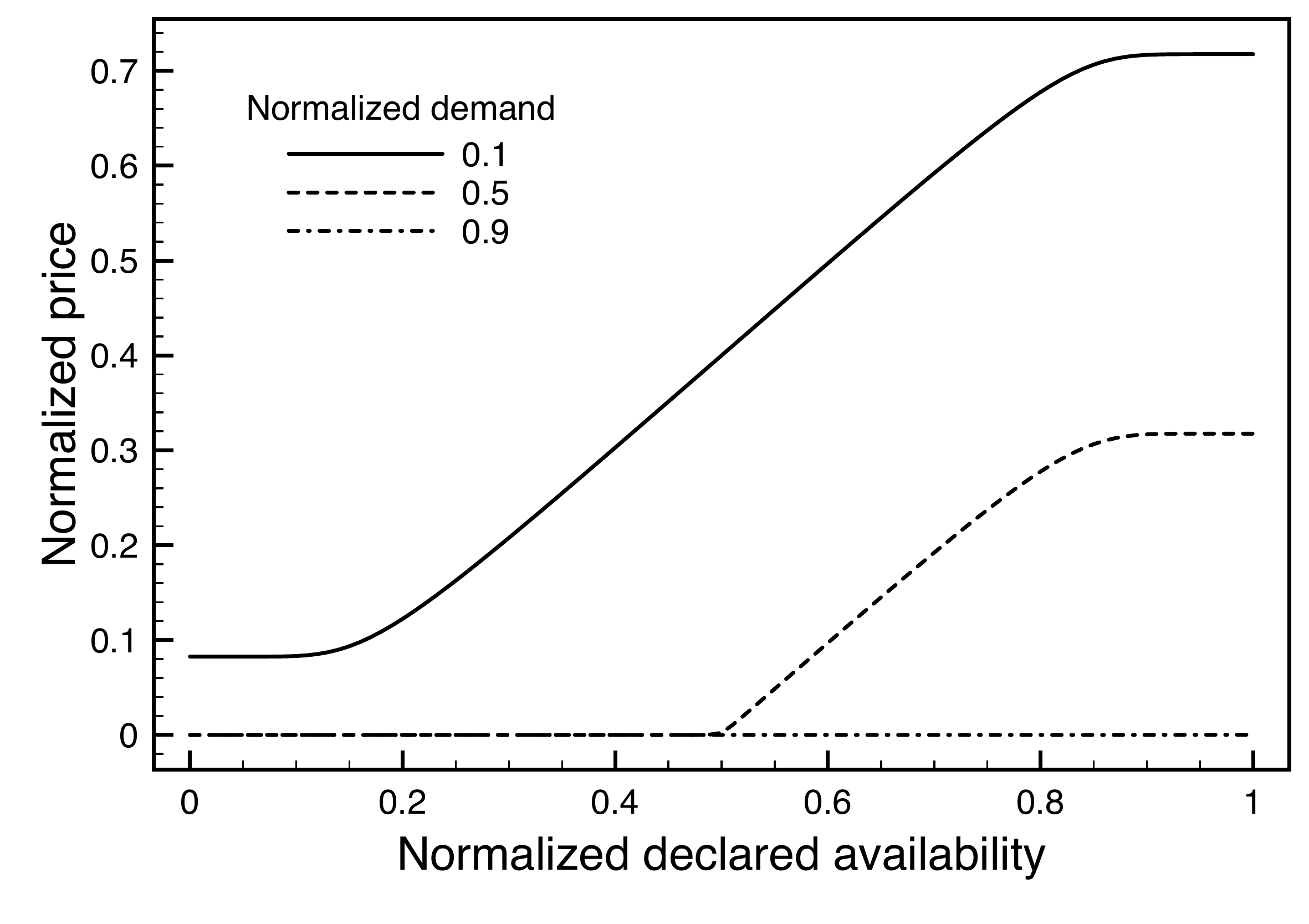} 
\caption{Impact of demand on option price (independent suppliers)}
\label{fig:prixind1}
\end{center}\end{figure}

We first plot the normalized option price $c_{\textrm{opt}}/c_{\textrm{s}}n$ for three different values of demand (again, with $n=100$ suppliers, $\lambda = 1.5$, and $p=0.5$) in \figurename~\ref{fig:prixind1}.
We see again the option price transitioning from a very low value to a high one as the suppliers declare a higher availability. From Equation (\ref{ind2}) the low and high value result
\begin{equation}
\begin{split}
\min c_{\textrm{opt}} &= c_{\textrm{s}} \frac{\sum_{i=k^{*}+1}^{n}(i-k^{*})\binom{n}{i} p^{i}(1-p)^{n-i} e^{-\lambda i }}{\sum_{j=0}^{n}\binom{n}{j}p^{j}(1-p)^{n-j} e^{-\lambda j }} \\
\max c_{\textrm{opt}} &= c_{\textrm{s}} \frac{\sum_{i=k^{*}+1}^{n}(i-k^{*})\binom{n}{i} p^{i}(1-p)^{n-i} e^{-\lambda (n-i) }}{\sum_{j=0}^{n}\binom{n}{j} p^{j}(1-p)^{n-j} e^{-\lambda (n-j) }} \\
&= c_{\textrm{s}} \frac{\sum_{i=k^{*}+1}^{n}(i-k^{*})\binom{n}{i} p^{i}(1-p)^{n-i} e^{\lambda i }}{\sum_{j=0}^{n}\binom{n}{j} p^{j}(1-p)^{n-j} e^{\lambda j }} \\\end{split}
\end{equation}
But there are two important differences with respect to the case of perfectly correlated suppliers:
\begin{itemize}
\item the low value is not always practically zero, but rises over zero when the demand is low;
\item the transition from low to high is quite smooth rather than abrupt as in the cases of perfect correlation.   
\end{itemize}

\begin{figure}
\begin{center}
\includegraphics[scale=0.4]{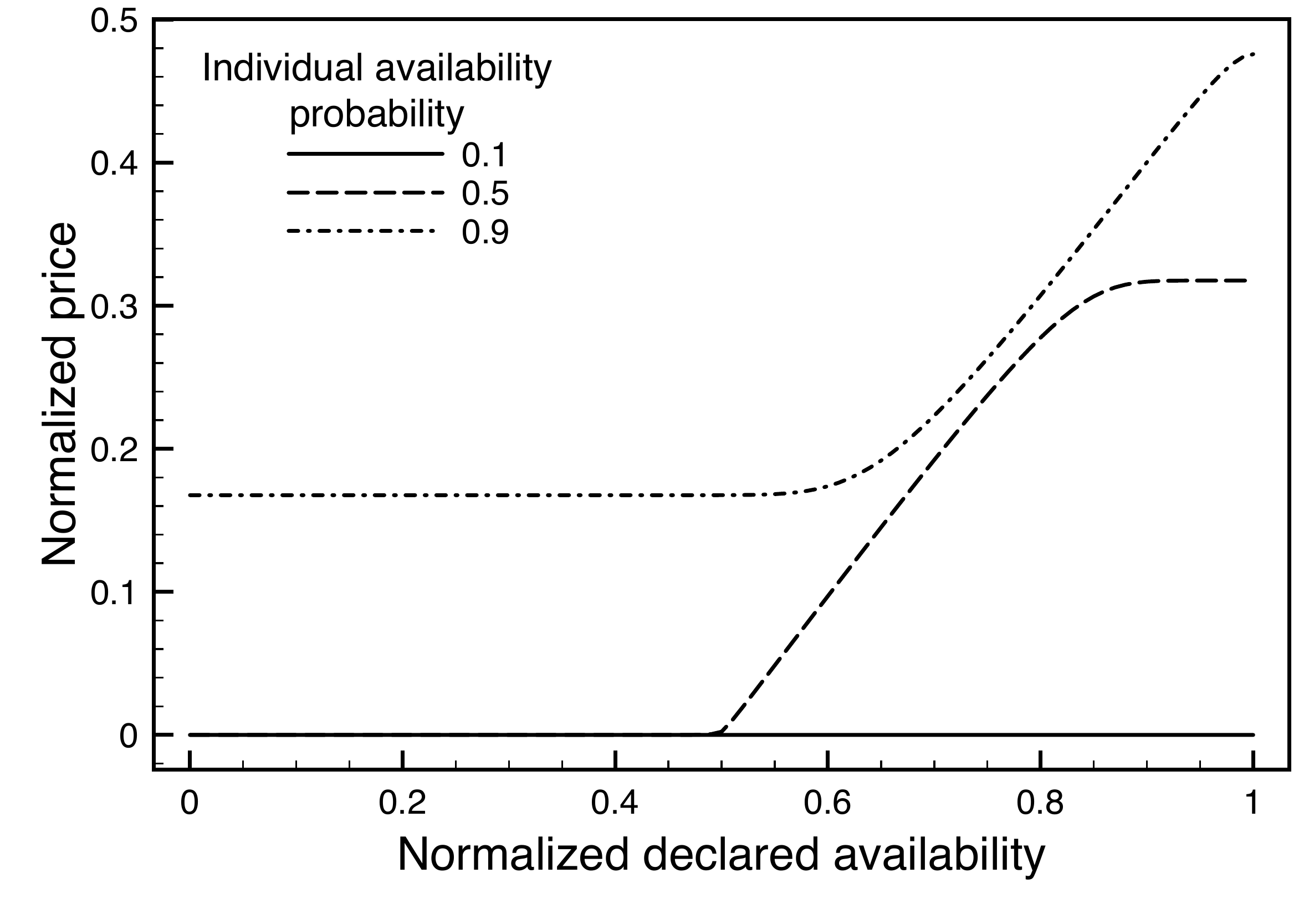} 
\caption{Impact of individual availability probability on option price (independent suppliers)}
\label{fig:prixind2}
\end{center}\end{figure}

In order to examine the impact of the probability of individual availability, we can consider the set of curves in \figurename~\ref{fig:prixind2}, where $n=100$, $k^{*}=50$, and $\lambda=1.5$. We observe a similar behaviour as in the previous curves, where now higher values of the individual probability push the price up.

Summing up, we can conclude that low demand, high declared availability, and high individual availability probability lead to higher option prices.

\subsection{Uniform availability}
The third model we consider is the uniform distribution, which is tantamount to assuming that we have no specific hypothesis for the availability of items.

As in  the previous two models, we proceed to define the extra-cost
\begin{equation}
\label{uni1}
\mathbb{E}[c_{\textrm{s}}(k-k^{*})^{+}\vert x<\hat{k}<x+dx] = \sum_{i=k^{*}+1}^{n} c_{\textrm{s}}(i-k^{*})\mathbb{P}[k=i\vert x<\hat{k}<x+dx],
\end{equation} 
and to evaluate the conditional probability through Bayes' theorem (in this case the uniform distribution is a non informative prior)
\begin{equation}
\begin{split}
\mathbb{P}[k=i \vert x<\hat{k}<x+dx] &= \frac{\mathbb{P}[k=i]\mathbb{P}[x<\hat{k}<x+dx \vert k=i]}{\mathbb{P}[x<\hat{k}<x+dx]}\\
&= \frac{\frac{1}{n+1}\frac{1}{2}\lambda e^{-\lambda\vert x-i \vert}}{\sum_{j=0}^{n}\frac{1}{n+1}\frac{1}{2}\lambda e^{-\lambda\vert x-j \vert}}\\
&= \frac{ e^{-\lambda\vert x-i \vert}}{\sum_{j=0}^{n} e^{-\lambda\vert x-j \vert}},
\end{split}
\end{equation}
which, replaced in Equation (\ref{uni1}), provides us with the final expression of the extra-cost
\begin{equation}
\label{uni2}
\begin{split}
c_{\textrm{opt}}  &= \sum_{i=k^{*}+1}^{n} c_{\textrm{s}}(i-k^{*})\frac{ e^{-\lambda\vert x-i \vert}}{\sum_{j=0}^{n} e^{-\lambda\vert x-j \vert}}\\
&= c_{\textrm{s}} \frac{\sum_{i=k^{*}+1}^{n}(i-k^{*}) e^{-\lambda\vert x-i \vert}}{\sum_{j=0}^{n} e^{-\lambda\vert x-j \vert}}
\end{split}
\end{equation} 

\begin{figure}
\begin{center}
\includegraphics[scale=0.4]{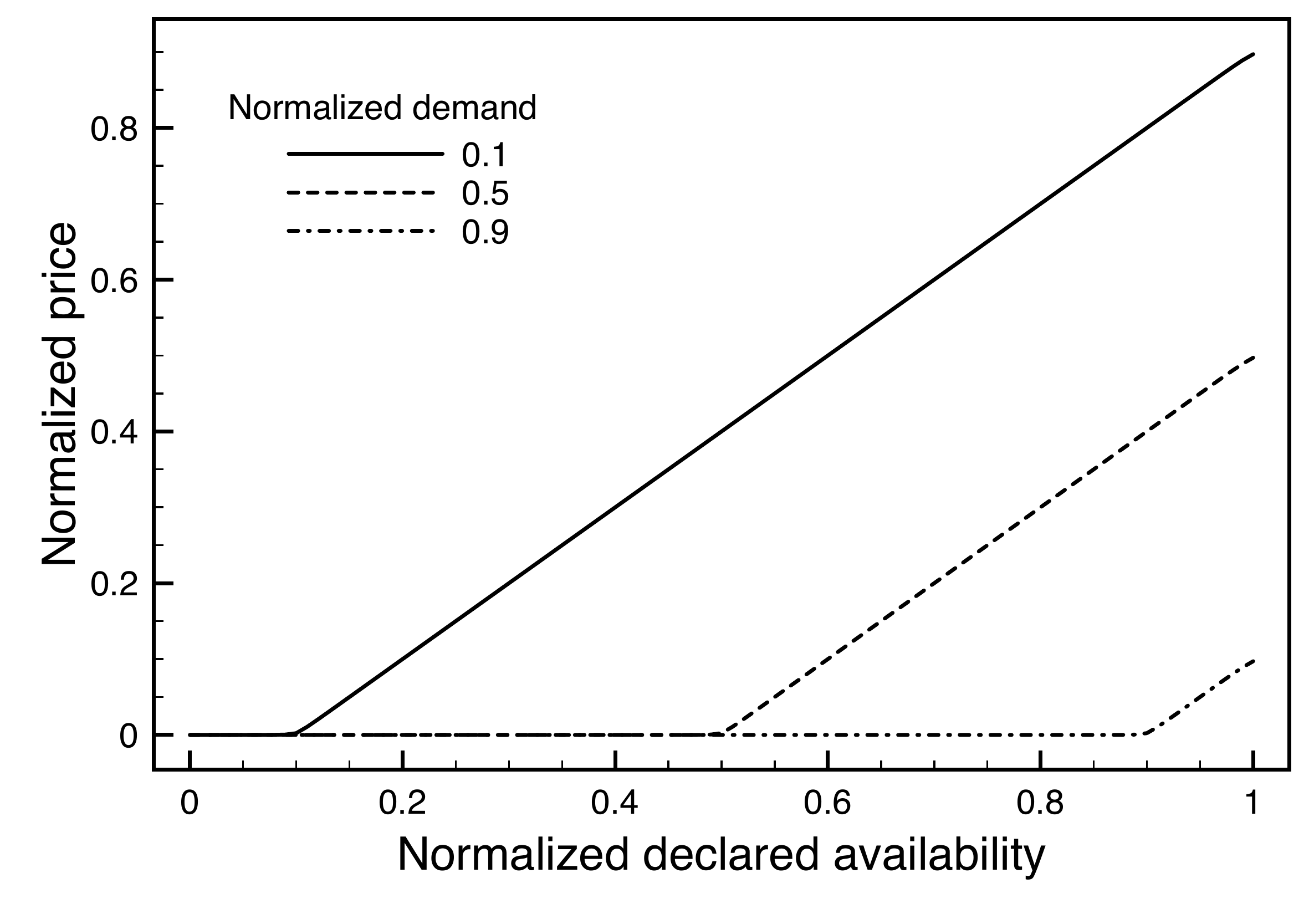} 
\caption{Impact of individual availability probability on option price (uniform availability)}
\label{fig:prixunif}
\end{center}\end{figure}

Now the only parameters are the demand $k^{*}$ and the declared availability $\hat{k}$. We plot three sample curves for the normalized price $c_{\textrm{opt}}/c_{\textrm{s}}n$ in \figurename~\ref{fig:prixunif}, again for $n=100$ suppliers and $\lambda = 1.5$.
We see that we obtain a piecewise linear curve with a knee in $\hat{k}=k^{*}$, that can be approximated by the formula
\begin{equation}
\frac{c_{\textrm{opt}}}{c_{\textrm{s}}n} = \frac{(\hat{k}-k^{*})^{+} }{n}.  
\end{equation}

In this case the risk transfer may be partial or excessive. If the noise added is negative, so that the suppliers declare an availability lower than real ($\hat{k}<k$), the price of the option is  
\begin{equation}
c_{\textrm{opt}} = c_{\textrm{s}}(\hat{k}-k^{*})^{+} < c_{\textrm{s}}(k-k^{*})^{+}, 
\end{equation}
so that not the whole risk is transferred to end customers. The opposite case occurs when $\hat{k}>k$, which may make end customers pay more than the actual risk.

\section{Conclusions}
The issue of option contracts in a privacy-aware market has been analysed, where the identity and level of stock of suppliers are kept hidden from end customers and potential competitors through a differential privacy scheme by the addition of Laplace noise. The scheme employs a broker that acts as an intermediary between suppliers and end customers. Pricing formulas for the option have been derived under three different models for the availability of items, which respectively assume a perfect correlation between suppliers, their independence (hence, perfect uncorrelation), or a uniform distribution (hence, a mild correlation). The option contract allows the broker to transfer part of its risk to end customers.

%
%

%
% ---- Bibliography ----
%
%\bibliographystyle{splncs03}
%\bibliography{Bib-privacy}  % sigproc.bib is the name of the Bibliography in this case

%\clearpage
%\addtocmark[2]{Author Index} % additional numbered TOC entry
%\renewcommand{\indexname}{Author Index}
%\printindex
%\clearpage
%\addtocmark[2]{Subject Index} % additional numbered TOC entry
%\markboth{Subject Index}{Subject Index}
%\renewcommand{\indexname}{Subject Index}
%\input{subjidx.ind}
\end{document}